\documentclass[10pt]{article}
\usepackage[utf8]{inputenc}
\usepackage[OE]{express}
\usepackage{braket}
\usepackage[alsoload=hep]{siunitx}
\usepackage{hyperref}
\usepackage{ae} %Computer Modern fonts. Not preferred, but otherwise doesn't compile on PVG's computer.

\begin{document}
\title{High-performance Raman memory with spatio-temporal reversal}

\author{Pierre Vernaz-Gris,\authormark{1,2} Aaron D. Tranter,\authormark{1} Jesse L. Everett,\authormark{1} Anthony C. Leung,\authormark{1} Karun V. Paul,\authormark{1} Geoff T. Campbell,\authormark{1} Ping Koy Lam,\authormark{1} and Ben C. Buchler\authormark{1,*}}

\address{\authormark{1}Centre for Quantum Computation and Communication Technology, Research School of Physics and Engineering, The Australian National University, Canberra, ACT 2601, Australia\\
	\authormark{2}Laboratoire Kastler Brossel, UPMC-Sorbonne Universit\'es, CNRS, ENS-PSL Research University, Coll\`ege de France, 4 place
	Jussieu, 75005 Paris, France}

\email{\authormark{*}ben.buchler@anu.edu.au}

	
\begin{abstract*}
A number of techniques exist to use an ensemble of atoms as a quantum memory for light. Many of these propose to use backward retrieval as a way to improve the storage and recall efficiency. We report on a demonstration of an off-resonant Raman memory that uses backward retrieval to achieve an efficiency of $65\pm6$\% at a storage time of one pulse duration. The memory has a characteristic decay time of 60 $\mu$s, corresponding to a delay-bandwidth product of $160$. %64/100 words
\end{abstract*}
	
%	\ocis{(190.4420) Nonlinear optics, transverse effects in;
%		(190.5650) Raman effect; (270.0270) Quantum optics; (270.1670)
%		Coherent optical effects; (270.5565) Quantum communications;
%		(270.5585) Quantum information and processing.}
	
%	\begin{thebibliography}{99}
%		
%		\bibitem{gallo99} K. Gallo and G. Assanto, ``All-optical diode based on second-harmonic generation in an asymmetric waveguide,'' \josab {\bfseries 16}(2), 267--269 (1999).
%		
%	\end{thebibliography}

\noindent \bibliographystyle{osajnl}
\bibliography{biblio_RamanMem}

\section{Introduction}

Developing quantum memories for light is an important challenge for implementing long-distance quantum networks \cite{Sangouard2011,Briegel1998} or optical quantum computers \cite{Nunn2013,Felinto2006}. To this end, a wide variety of memory platforms and protocols have been developed to store and retrieve photons or other quantum states of light. Many of these rely on absorbing or scattering the information carried by the light into an ensemble of atoms in such a way that it can be converted back into light at a later time. One of the difficulties to doing so efficiently is that, once retrieved, the light may be re-absorbed by atoms in the ensemble. To avoid re-absorption, most quantum memory protocols use either backward retrieval \cite{Afzelius2009,Kozhekin2000,Moiseev2001} or a spatial inversion of the ensemble \cite{Hetet2008,Longdell2008} to exploit a time-reversal symmetry in the dynamics of the light-atom system \cite{Gorshkov2007b,Nunn2007,LeGouet2009,Moiseev2011} . While higher recall efficiencies can in principle be achieved using backward retrieval, most experimental demonstrations of memory protocols use forward retrieval for simplicity.

One protocol that shows particular promise for the storage of high-bandwidth quantum information is the off-resonant Raman memory. This technique uses a strong, pulsed optical control field to coherently scatter a weaker signal field, encoding the information carried by the signal into a long-lived spin state of the atomic ensemble. A second control field pulse scatters the stored information back into an optical signal. This approach with forward retrieval has been used to demonstrate single-photon \cite{Reim2011} , high-fidelity \cite{England2012}, or multimode \cite{Nunn2008} memories in warm atomic vapour, quantum memory in hollow-core fibres \cite{Sprague2014}, and terahertz-bandwidth storage in molecules \cite{Bustard2013} and diamond \cite{Lee2012}. Raman memory has also enabled the storage of polarisation entanglement in cold atomic ensembles \cite{Ding2015}. Specifically in this protocol, backward retrieval maximises the use of the optical depth of the ensemble on the storage and retrieval steps, whereas forward-retrieval schemes necessarily compromise on each of these to maximise the overall efficiency \cite{Gorshkov2007b}. %also more details spinwave shape?
%Using a backward-propagating control pulse retrieves the signal in the backward direction, enabling a high efficiency due to the time-reversal symmetry of the storage and retrieval processes.

Here we report the first demonstration of an off-resonant Raman memory that uses backward retrieval. We implement the memory using a cloud of cold atoms with an optical depth (OD) of 300. We achieved a recall efficiency of $65\pm6\%$ and, under different conditions, a delay-bandwidth product of $160$. It is the first reported Raman memory with an efficiency above $50\%$, which is an important criterion for unconditional security in quantum communication protocols \cite{Grosshans2001}.

\section{Atomic ensemble preparation}

\paragraph{Preparation of cold atoms}

\begin{figure}[htb]
\centering{}\includegraphics[width=0.7\linewidth]{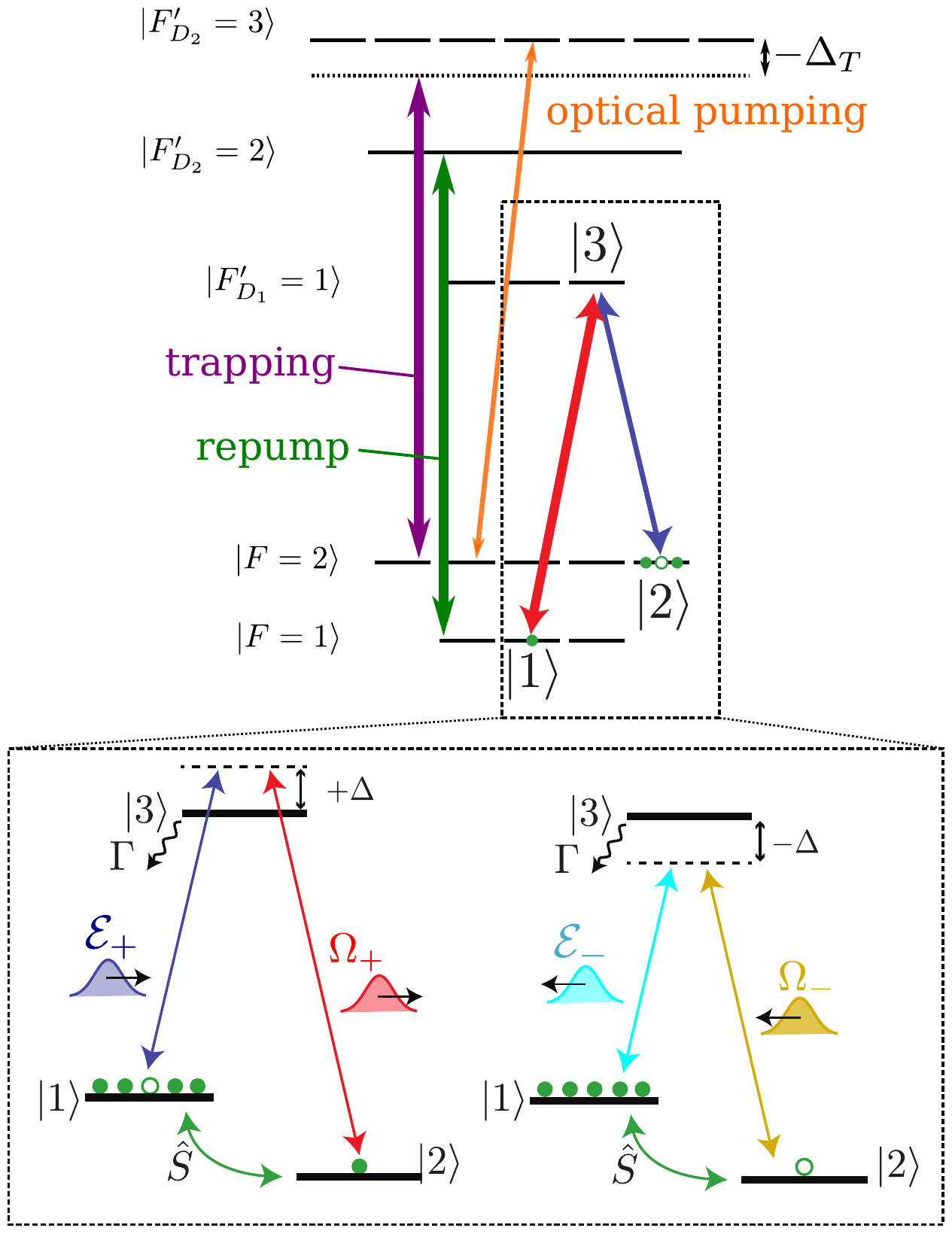}
\caption{The atomic level configuration. Trapping is performed with light that is red-detuned from the $\ket{5S_{1/2},F=2}\to\ket{5P_{3/2},F'=3}$ transition and a repump is applied on the $\ket{5S_{1/2},F=1}\to\ket{5P_{3/2},F'=2}$ transition. After trapping and compression of the ensemble, a $\sigma^+$-polarized optical pumping beam is applied to the $\ket{5S_{1/2},F=1}\to\ket{5P_{3/2},F'=2}$ transition in conjunction with the repump to populate the $\ket{5S_{1/2},F=2, m_f=+2}$ state. Inset: Left: A signal $\mathcal{E}_{+}$ traverses an elongated atomic cloud where a two-photon resonant control beam $\Omega_{+}$ converts it to a collective atomic excitation. Right: A backward-propagating control beam $\Omega_{-}$ subsequently retrieves the signal in the backward direction, as $\mathcal{E}_{-}$. \label{fig:levelscheme}}
\end{figure}

A $\SI 5{\centi\meter}$-long cloud of rubidium-87 atoms is prepared using an elongated magneto-optical trap with dynamic control of the magnetic gradient, cooling beam frequency and repump intensity.

\begin{figure}[htbp]
	\centering{}\includegraphics[width=0.95\linewidth]{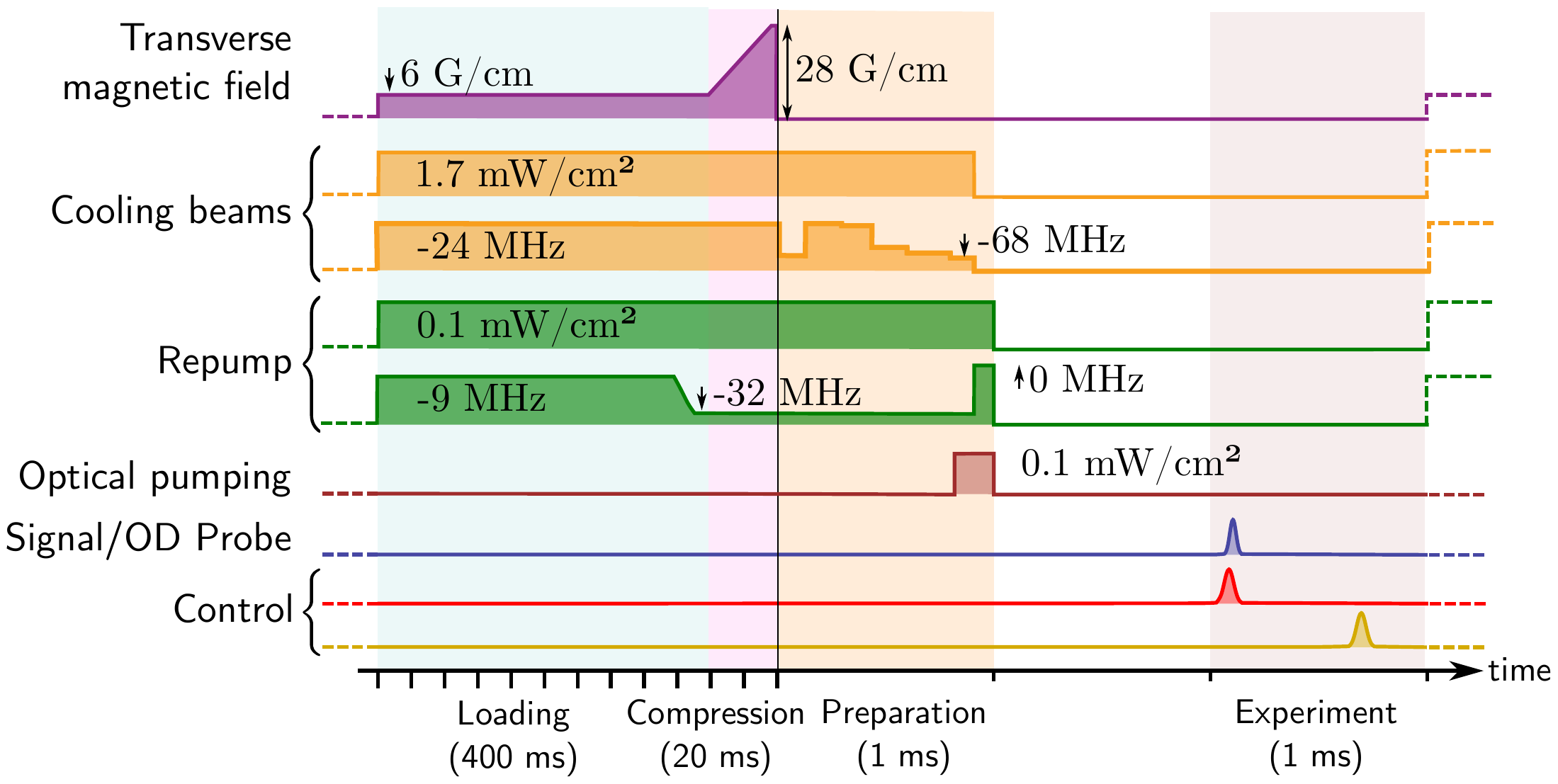}
	\caption{The atomic preparation sequence. The compression and polarisation-gradient cooling sequences are divided into 20 and 5 time bins, respectively, and are passed onto a machine-learning algorithm. This algorithm determines the optimal values of repump and cooling beam frequencies, and current through the transverse coils for magnetic trapping. A typical optimised set of parameters is presented here.\label{fig:sequence} }
\end{figure}

The two-dimensional magneto-optical trap consists of a set of coils for radial confinement with a magnetic gradient of $\SI 6{\gauss\per\centi\meter}$, completed by a pair of coils for longitudinal capping, and three retro-reflected two-inch-diameter cooling beams that intersect at the zero-magnetic-field location. The atomic transitions involved in the ensemble preparation is shown in Fig.~\ref{fig:levelscheme}. A $\SI{20}{\milli\second}$ sequence further compresses and cools the ensemble by employing the temporal dark spontaneous-force optical trapping (SPOT) technique \cite{Ketterle1993}. Simultaneously, the current in the radial confinement coils is increased to reduce the size of the trap and compress the atoms, thereby increasing the axial optical depth. The Earth's magnetic field is cancelled by a set of three orthogonal 1-meter-diameter coils and a uniform bias magnetic field of $0.5$ G is applied along the optical axis direction to lift the degeneracy between the Zeeman sub-levels and the atoms are transferred to the $\ket 1=\ket{5S_{1/2},F=2,m_{F}=+2}$ edge state using a $\SI{\sim0.7}{\milli\watt\per\centi\meter\squared}$ $\sigma_{+}$-polarised optical pumping beam at the same frequency as the cooling beam, along the quantisation axis. See Fig.~\ref{fig:sequence} for the detailed sequence of the ensemble preparation parameters.

%\begin{figure*}[htbp]
%\end{figure*}

\paragraph{Machine-learned compression-and-cooling sequence}

For an enhanced compression, cooling and optical pumping sequence, control over the time profile of the cooling frequency, repump frequency and magnetic gradient is given to a machine-learning algorithm. The ensemble compression phase is divided into 20 time bins of $\SI 1{\milli\second}$. The values of the repump frequency and magnetic gradient at each time bin are parameterised to linear ramps only. During the temporal dark SPOT the repump frequency is red-detuned from resonance by $\SI{32}{\mega\hertz}$. The trapping frequency is divided into 5 time bins over a period of $\SI 1{\milli\second}$ during a polarisation-gradient cooling (PGC) and optical pumping phase, which we call the preparation phase. The values at these points are not restricted to a ramp and can be changed independently. %Figure \ref{fig:ML} shows a typical machine-optimised sequence.

The machine-learning optimisation is based on a learner which uses a Gaussian process to infer a statistical model \cite{Hush2016}. A similar method has recently been used to optimise the evaporative cooling of a Bose-Einstein condensate \cite{Wigley2016}, but also dynamical decoupling \cite{Mavadia2017} and feedback control of a qubit \cite{August2017}.

A probe pulse close to resonance on the $\ket 1\to\ket{5P_{1/2},F'=1}$ transition replaces the signal in the experiment sequence, shown in Fig.~\ref{fig:sequence}. The absorption of this probe is monitored by a detector (Thorlabs APD120A) placed in the beam path after the atoms, which we call the forward detector, whose signal is used as feedback for the machine-learning algorithm. The amount of absorption is linked to the OD. We define transmission to be the cost function, which is actively minimised.

The experimental conditions vary noticeably over days. The optimal parameters for cooling and compression are investigated once a day, before each memory run. A typical profile of the resulting compression and preparation sequence is presented in Fig.~\ref{fig:sequence}. 
%With the optimised sequence, we can consistently obtain ODs of the order of 300, a $75\%$ improvement compared to our previous compression ramp profile and PGC sequence.

\section{Backward-Raman memory}
\paragraph{Memory scheme, relevant transitions}

\begin{figure}[htbp]
	\centering{}\includegraphics[width=0.7\linewidth]{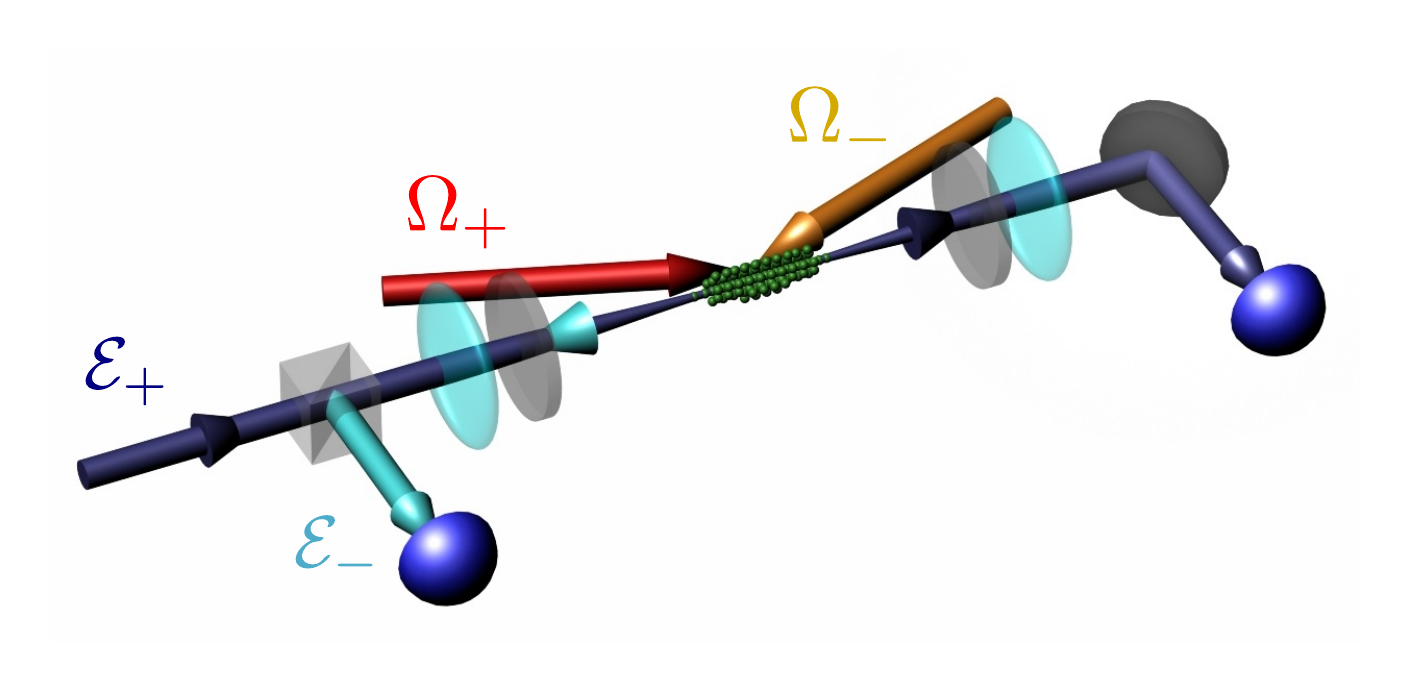}
	\caption{The signal $\mathcal{E}_{+}$ traverses the atomic cloud where a two-photon resonant control beam $\Omega_+$ converts it to a collective atomic excitation. A reference without atoms or the fraction which is not absorbed is monitored on the right-hand detector. A backward-propagating control  $\Omega_{-}$ subsequently retrieves the signal in the backward direction, as $\mathcal{E}_{-}$ which is sent to the left-hand detector. \label{fig:Ramanscheme} }
\end{figure}

The Raman memory scheme consists of converting an incoming signal light pulse into a spin-wave between the two hyperfine ground levels of rubidium 87: $\ket 1=\ket{5S_{1/2},F=2,m_{F}=+2}$ and $\ket 2=\ket{5S_{1/2},F=1,m_{F}=0}$. The signal, denoted $\mathcal{E}_{+}$, is blue-detuned by $\SI{230}{\mega\hertz}$ from the transition $\ket 1\to\ket 3=\ket{5P_{1/2},F'=1,m_{F'}=+1}$. A forward-propagating control, denoted $\Omega_{+}$, equally detuned from $\ket 2\to\ket 3$, forms an angle of $\approx\SI 6{\milli\radian}$ with the signal. The $\Omega_{+}$ control maps the signal onto a spin-wave. A symmetrically(red)-detuned control, $\Omega_{-}$, is a mirror image of $\Omega_{+}$. $\Omega_{-}$ addresses the spin-wave and couples it to the $\SI{230}{\mega\hertz}$-red-detuned signal, noted $\mathcal{E}_{-}$, which is the counter-propagating version of $\mathcal{E}_{+}$ in the same spatial mode (see Fig.~\ref{fig:Ramanscheme}).

Control and signal laser light is provided by a titanium-sapphire laser (MSquared SolsTiS) which is frequency-locked to the saturated absorption of the $\ket 2\to\ket{5P_{1/2},F'=1}$ transition of rubidium-87 around $\SI{795}{\nano\meter}$. Light resonant with the $\ket 1\to\ket{5P_{1/2},F'=1}$ transition is obtained from filtering out the $+\SI{6.8}{\giga\hertz}$ sideband of an EOM-phase-modulated beam. Acousto-optical modulators enable independent temporal pulse shaping of the signal and the forward and backward control beams.

\subsection{Storage-and-retrieval efficiency}

A Gaussian-shaped signal pulse with a full width at half-maximum of $\SI{5}{\micro\second}$ is coherently mapped onto the atomic ensemble and subsequently recalled in the backward direction with a pair of forward and backward Gaussian-shaped control pulses. The width, delay, and intensity of the forward control pulse for storage is optimised to minimise the leakage of signal light through the atoms. We found that the optimal storage control pulse had a width of $\SI 9{\micro\second}$ and preceded the signal pulse by $\SI{5}{\micro\second}$. The retrieval control pulse, backward-propagating, is given the same shape as the optimal storage pulse, and with a larger intensity \cite{Nunn2007}.

The memory efficiency is defined as the ratio between the integration of the retrieved signal on the backward detector and the signal without atoms on the forward detector, corrected for the detection losses and sensitivities. The highest reproducible memory efficiency after a storage time of one pulse width is found to be $(65\pm6)\%$. The large uncertainty is owed to the necessary imperfection of the measurement of the optical losses on the two different detection paths.

This figure more than doubles the previous recorded efficiency of $30\%$ for an optical Raman memory and also outperforms the maximum achievable efficiency of $54\%$ with Controlled-Reversed Inhomogeneous Broadening with forward retrieval \cite{Sangouard2007,Longdell2008}. Although Raman memories do not have an efficiency limit for forward retrieval \cite{Nunn2007}, the time-reversal symmetry of backward retrieval is predicted to yield a higher efficiency if the optical depth is held constant.

Contrary to previous Raman memory experiments operating at very large detuning with ultra-short pulses \cite{Nunn2008,Surmacz2008,Nunn2007,Michelberger2015,Saunders2016,Reim2012,Reim2010}, the signal can be incoherently absorbed by the excited level, which fundamentally limits the memory efficiency. At an OD of $500$, we note that this incoherent absorption accounts for more than $5\%$ loss, which is experimentally verified by comparing the forward detector signal with and without atoms.

\subsection{Memory lifetime}

\begin{figure}[htbp]
\centering \includegraphics[width=\linewidth]{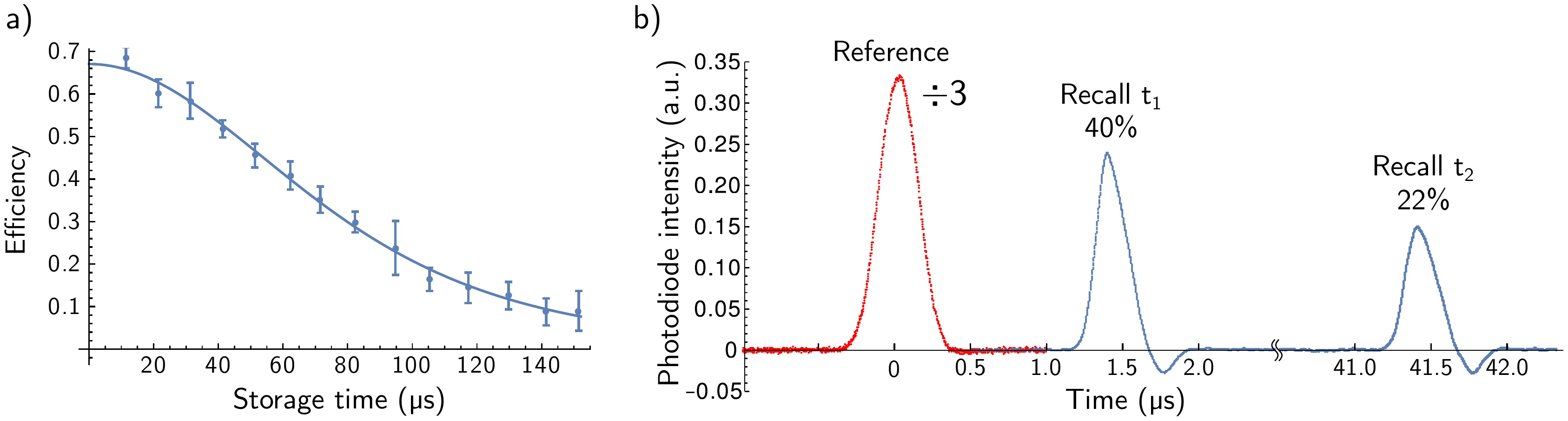}

\caption{a) Decay of the memory efficiency with storage time. The error bars correspond to the standard deviation. The efficiency is fitted to a motional dephasing model with an initial efficiency $\eta_{0}=(69\pm6)\%$ and diffusion and transit times $\tau_{D}=\protect\SI{110}{\micro\second}$ and $\tau_{T}=\protect\SI{170}{\micro\second}$, respectively. b) Storage and recall of a $\SI{360}{\nano\second}$ $1/e^{2}$-width pulse. Photodiode signal for reference (scaled down) and recalls at storage times $t_1=\SI{1.5}{\micro\second}$ and $t_2=\SI{41}{\micro\second}$ are shown. The characteristic decay time corresponds to a delay-bandwidth product of $160$.\label{fig:lifetime}}

\end{figure}

The storage duration is determined by the delay of the read-out control beam, which can be controlled at will. The memory efficiency is measured after different storage durations, shown in Fig.~\ref{fig:lifetime}. The memory efficiency over time is fitted by the motional dephasing model described in \cite{Jenkins2012}, which derives an efficiency of the form:

\[
\eta(t)=\frac{\eta_{0}}{\left(1+(t/\tau_{D})^{2}\right)^{2}}\exp\left[-\frac{(t/\tau_{T})^{2}}{1+(t/\tau_{D})^{2}}\right]
\]

with efficiency at zero storage duration $\eta_{0}=69\pm6\%$, diffusion and transit characteristic times $\tau_{D}=\SI{110}{\micro\second}$ and $\tau_{T}=\SI{170}{\micro\second}$. Figure \ref{fig:lifetime}(a)
shows the experimental efficiency measurement results. 

\subsection{Delay-bandwidth product}

We have employed the same protocol for the storage of shorter pulses, demonstrating large delay-bandwidth products with little compromise on the efficiency. To maintain a comparable efficiency, the control beams intensity had to be increased for shorter signal pulse widths and the memory was eventually limited by the available laser power, as well as the optical depth of the ensemble. The highest delay-bandwidth product of $160$ was obtained for a signal with a $\SI{360}{\nano\second}$ $1/e^{2}$-width, a storage efficiency after one pulse-width of $40\%$ and a characteristic decay time $\tau=\SI{60}{\micro\second}$. Figure \ref{fig:lifetime}(b) shows the reference experimental photodiode signal for the  pulse to be stored in the memory, and the signals for recalls at times $t_1=\SI{1.5}{\micro\second}$ and $t_2=\SI{41}{\micro\second}$. %The ring-down at the end of the pulse is from operating close to the photodiode bandwidth (but not on the reference...)

\subsection{Beam-splitting results}

\begin{figure}[ht]
\centering \includegraphics[width=0.55\linewidth]{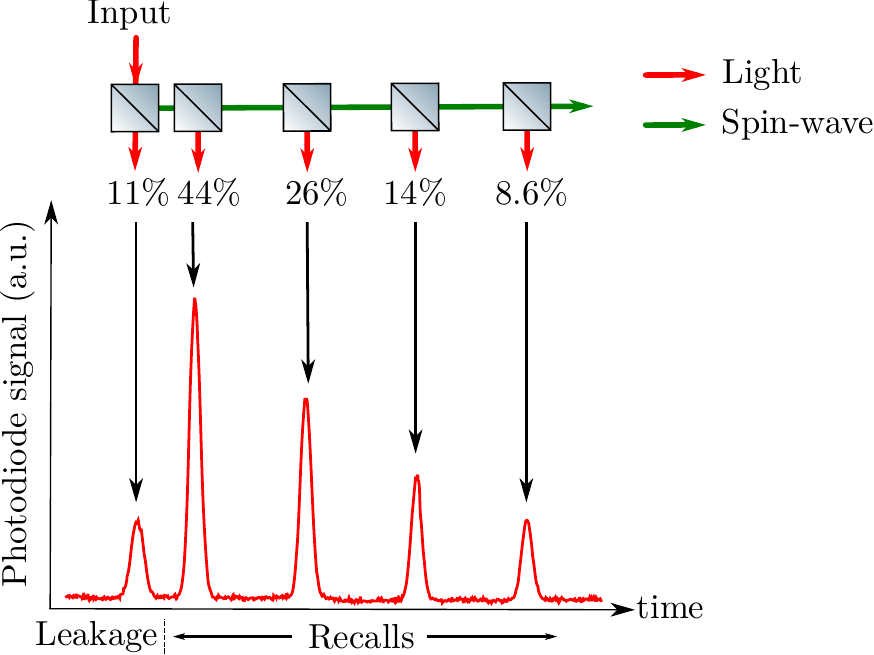}
\caption{Schematic of an equivalent beam-splitter array. The input light is partially converted to a spin-wave and the rest leaks through the atomic ensemble and is detected on the photodiode.  It is estimated about 35\% of the light that is absorbed is not mode-matched to be efficiently recalled, in this short-pulse configuration. At the given control beam intensity, each retrieval pulse extracts only about $60\pm4\%$ of the stored spin-wave, the remainder is left for a subsequent retrieval step. The memory decay between retrievals is ignored at this time scale.\label{fig:beamsplitter}}
 
\end{figure}

The implemented Raman memory can also be used as dynamically-reprogrammable array of beam-splitters, an application first demonstrated in Raman memories in \cite{Reim2012} and depicted in Fig.~\ref{fig:beamsplitter}. The control power was chosen weak enough to read out only about $60\%$ of the remaining atomic spin-wave with each recall pulse. This beam-splitting operation enables interference between the memory read-out and another input \cite{Campbell2012,Pinel2015}. This scheme of repeated weak recalls has also been shown to enable a doubly-exponential enhancement on the scaling of continuous-variable entanglement distillation \cite{Datta2012}.

\section{Discussion}

Phase-matching between the spin-wave and the optical fields is known to determine the direction of the retrieved signal \cite{Braje2004} and has also been shown to impose a limit on the achievable efficiency of memories in forward- and backward-retrieval configurations \cite{Surmacz2008}. The memory efficiency benefits from the backward-retrieval geometry of our realisation, as it more than doubles the record storage-and-recall efficiency for a Raman memory \cite{Reim2011}. We have also found that the memory lifetime is limited by atomic motion, which causes the blurring of the spin-wave. This had been predicted as a main limitation on the backward-retrieval Raman memory scheme \cite{Tikhonov2015,Surmacz2008}.

%We note, however, that in this work we do not characterise the noise performance of the memory to verify that we have surpassed the no-cloning limit. The experimental scheme can be modified to store single-photon level coherent beams or heralded sub-megahertz-linewidth single photons from pairs of photons generated from four-wave mixing in another atomic ensemble \cite{Du2008} or triply-resonant spontaneous parametric down-conversion \cite{Rambach2016}. Stable frequency-dependent isolation from the control while maintaining high signal transmission can be achieved by adding monolithic filter cavities before the detectors \cite{Palittapongarnpim2012,Ahlrichs2013}.

Few experiments employ elements of a symmetrical geometry in Raman systems, although this symmetry is found to enhance the conversion of light to spin-waves \cite{Chen2013a,Guo2017}. To our knowledge, however, our realisation is the first Raman memory with backward retrieval.

\section*{Funding}

The Australian Research Council (ARC) (CE110001027, FL150100019, FT100100048).

\section*{Acknowledgments}

We would like to thank Nicholas P. Robins for the cold atom setup and Michael R. Hush for help with the machine-learning algorithm. 

\end{document}